\documentclass[aps,prl,reprint,groupedaddress]{revtex4-1}

\usepackage{graphicx}

\begin{document}


\title{Electron Interaction Effects in Periodic Driven Kitaev Model: Topology Breaking and Enhancement of Quantum Chaos}


\author{W. Su}
\author{M. N. Chen}
\author{L. B. Shao}
\author{L. Sheng}
\author{D. Y. Xing}

\affiliation{National Laboratory of Solid State Microstructures, School of Physics, and
Collaborative Innovation Center of Advanced Microstructures, Nanjing University, Nanjing 210093, China}


\date{\today}

\begin{abstract}
The effect of electron-electron interaction on  Floquet topological superconducting chains  is investigated
numerically through full diagonalization and time evolution. The preservation of topology in the weak interacting regime is represented by a many-body form of the Majorana survival probability, and the emergence of chaos is characterized using the level statistics.  In the presence of weak interaction, there
appear a multitude of avoided crossings in quasi-energy spectra, and the resulting chaos is not full but
can coexist with the topology. Strong interaction will lead the system into a topologically trivial and fully chaotic phase.

\end{abstract}


\maketitle

\emph{Introduction.-}Topological insulators induced dynamically by time-periodic driving have been in a hot research in recent years\cite{rechtsman_photonic_2013,itin_effective_2015,lindner_floquet_2011,gomez-leon_floquet-bloch_2013,ho_quantized_2012,kitagawa_topological_2010}.
Quite often, the phase diagram in the Floquet system is remarkably richer than its stationary counterpart.
The periodic driving one-dimensional (1D) $p$-wave superconductor is an example of Floquet topological insulator\cite{thakurathi_floquet_2013,tong_generating_2013}.
It possesses two kinds of Floquet Majorana fermions (FMFs) for a long open chain, i.e., FMFs with zero quasi-energy and $\pi$ quasi-energy.
Those zero ($\pi$) modes share most properties of their stationary counterparts, i.e., non-Abelian braiding statistics and immunity to local decoherence.
Various promising proposals exist for creating FMFs in superconducting devices and cold-atom quantum wires\cite{liu_floquet_2013,jiang_majorana_2011,foster_quench-induced_2014}.

Studies on systems supporting FMFs  focused mostly on simple Hamiltonians within the single-particle framework.
While the influence of interaction on stationary systems that support Majorana fermions has been widely investigated both analytically and numerically\cite{katsura_exact_2015,stoudenmire_interaction_2011,ghazaryan_long-range_2015,gangadharaiah_majorana_2011,sela_majorana_2011,Hassler2012},
seldom existed for periodic driving ones. For FMF systems, the importance of interaction comes in two aspects.
First, it has influence on the topological phase in quantitative and perhaps even qualitative ways.
It is believed that interaction which is much smaller than the quasi-energy gap will not destroy the topological property of the FMF system.
However, what happens when the interaction strength is comparable to the quasi-energy gap?
Can the topological non-trivial phase still preserve in the strong interacting regime? (see \cite{sela_majorana_2011,Hassler2012}
for stationary topological superconductors with strong interaction)
Second, due to the folding structure of the Floquet spectrum, its excited state spectrum is of great importance, while in stationary systems
one mainly concern with the ground states (GS). Interaction combined with the periodic driving introduces non-integrability and makes the system into the phase of quantum chaos\cite{jacquod_emergence_1997,santos_onset_2010}.
Chaotic phenomenons are commonly occurred in Floquet systems with interaction, and is of particularly importance when the strength of interaction is comparable to the frequency of the driving period.
Chaotic and topological phenomena have opposite characteristics, the former being sensitive to disturbances and the latter exhibiting robustness.
A simple question then arises: whether or not the  topology and quantum chaos can coexsist in a certain parameter region?

The goal of this paper is to address the questions above.
We use a periodic driving superconducting chain of spinless fermions as a concrete realization.
In intermediate frequency region, the analytic procedures based on high-frequency\cite{itin_effective_2015} and low-frequency expansions are invalid.
We numerically use full diagonalization and time evolution procedure to investigate the topological properties of a finite chain, while the emergence of quantum chaos are demonstrated by level statistics.
The study of FMFs in the interacting systems will be in favor of understanding the stability of FMFs in real systems, which is a key property to realize topological quantum computing.
Moreover, it will be in favour of understanding how topology competes with quantum chaos.

\emph{Model and its symmetry.-}Consider a periodic driven 1D  spinless $p$-wave superconducting chain with $N$ lattice sites
in the presence of electron-electron interaction. The Hamiltonian is given by
\begin{eqnarray}
H(t)&=&\sum_{j=1}^{N-1}\left( -t_0 c_j^\dag c_{j+1} +\Delta c_j c_{j+1} +H.c. \right)\nonumber\\
&&+V\sum_{j=1}^{N-1}\left( \hat{n}_j-\frac{1}{2} \right)\left( \hat{n}_{j+1}-\frac{1}{2} \right)\nonumber\\
&&-\mu(t)\sum_{j=1}^{N}\left( \hat{n}_j-\frac{1}{2} \right).
\label{eq:2.1}
\end{eqnarray}
Here, $t_0$ is the hopping amplitude,
$\Delta$ is the $p$-wave superconducting order parameter,
$\mu$ is the time-periodic driving potential, and $V>0$ is the strength of the repulsive electron-electron interaction.
We set both the Planck's constant $\hbar$ and lattice constant to be unity.
For numerical convenience, the time-dependent chemical potential is taken to be square waves
\begin{equation}
  \mu(t)=\left\{
    \begin{array}{ll}
    \mu_0+\delta\mu,& nT<t<(n+1/2)T;\\
    \mu_0-\delta\mu,& (n+1/2)T<t<(n+1)T,
  \end{array}
\right.
  \label{eq:2.2}
\end{equation}
where $n=1,2,\cdots$, and $T$ is the driving period.
The evolution operator of a full time period, also known as
the Floquet operator, reads $\hat{U}(T,0)=e^{-iH_2 T/2}e^{-iH_1 T/2}$, with $H_1$ ($H_2$) the Hamiltonian defined in the graded Hilbert space for the first (second) half of a period.

The Hamiltonian
$H$ and thus $U$ respect the time-reversal symmetry and the
parity of the fermion number, which is defined as the fermion number modulo $2$.
Furthermore, the Hamiltonian has inversion symmetry $P$, which is the space reflection followed by a gauge transformation of $\Delta\to -\Delta$. Assuming $P^2=1$, one can find that
under the periodic boundary condition (PBC), only the subspace $k=0$ for an odd number of sites or
the subspace of $k=0$ and $k=\pi$ for an even number of sites can be further decomposed into subspaces of $P=\pm 1$, where $k$ is the total momentum. In the particular case of $\mu_0 = 0$,
the Floquet operator is invariant under charge conjugation $c_j\to (-1)^j c_j^\dagger$.

\begin{figure}[t]
\centering
\includegraphics[width=0.9\linewidth]{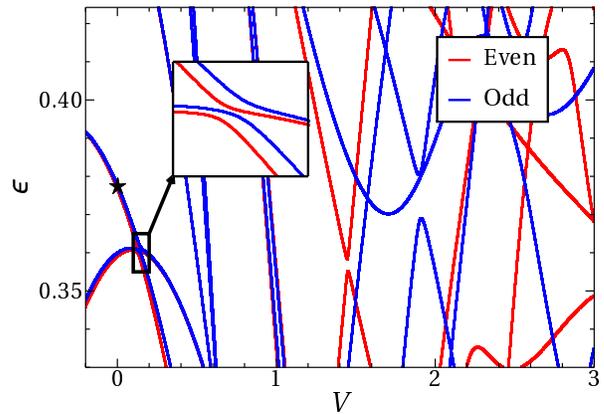}
\caption{Quasi-energy spectra of an open chain with even and odd fermion numbers in a range of interacting strength $V$. The spectra of even and odd sectors
are almost identical to each other for $V \lesssim 1$, suggesting a topologically nontrivial phase. The five-pointed star denotes the ground-state
quasi-energy in the interacting-free case. The other parameters are set to $L=8$, $\Delta=1$, $\mu_0=0$, $\delta\mu=2.5$, and $T=1$. }
\label{fig:spectrum}
\end{figure}

\emph{Quasi-energy spectra.-}
Fig. \ref{fig:spectrum} shows the quasi-energy spectra in a system with open boundary condition (OBC) for
$V<3$ with $t_0$ as the unit of energy, and the system is in the topologically nontrivial phase at $V=0$. The quasi-energy spectra are calculated by exactly diagonalizing  $U(T,0)$ for
even and odd fermion numbers separately.
In the stationary state, the difference in the GS energy between the even and odd fermion parity sectors serves as an order parameter to distinguish between the topologically trivial and nontrivial phases \cite{stoudenmire_interaction_2011,ghazaryan_long-range_2015}.
For a Floquet system, however, the concept of GS is not directly applicable due to its non-equilibrium nature.
Since Majorana fermions are modes with zero (or $\pi$) quasi-energy, which are localized at the ends of a long open chain,
it is expected that the degeneracy of the odd and even sectors is present for all the eigen-states.
Fig.\ref{fig:spectrum} shows these degeneracies for small $V$, corresponding to the topologically nontrival phase. For relatively larger $V$, the degeneracy, thus the topology, is destroyed.

In Fig. \ref{fig:spectrum}, one can see a multitude of avoided crossings, especially for $V>1$, where the system is in the topologically trivial phase. These avoided crossings are a signature of quantum chaos \cite{Haake2010}. In the case of $V<1$, the avoid crossings still exist, but their gaps are too narrow to be distinguished by naked eye.

\emph{Quenching dynamics.-}The difficultiy of extracting topological information through quasi-energy spectra comes from two aspects.
First, $U(T,0)$ is generally a dense matrix, although $H_1$ and $H_2$ are sparse matrices themselves.
Second, the quasi-energy is only defined in an interval of $[-\pi/T,\pi/T]$. In the thermodynamic limit, there are continuum spectra for the odd and even fermion parity sectors, and thus the degeneracy is obscure.

In the absence of the interaction ($V=0$), an effective Hamiltonian $h_f$ can be obtained as   $h_f=\sum_{j=1}^{L}\theta_j(f_j^\dag f_j-\frac{1}{2})$,~\cite{thakurathi_floquet_2013,Suppl}
where $f_j$ ($f_j^\dag$) is the Dirac fermion operator of the quasi-particle. The single-particle quasi-energies $\theta_j\in [0,\pi]$ have been arranged in an increasing order.
We define the ``GS'' as state $|0\rangle$ without any quasi-particles, i.e., $f_j|0\rangle=0$ for $j=1,\cdots,L$.
For the topological non-trivial phase, the other degenerated ``GS'' is $|1\rangle\equiv f_1^\dag|0\rangle$ (or $|L\rangle\equiv f_L^\dag|0\rangle$ for the existence of $\pi$ mode), which has a quasi-particle with zero (or $\pi$) quasi-energy.

Starting from a topologically non-trivial phase at $V=0$, we slowly increment $V$ sequentially each period, i.e., $V(n_p)=n_p V_M/N_T$. Here $n_p$ is the number of driving period, $V_M$ is the final $V$, and the inverse of the total number of period $1/N_T$ serves as the rate of change. The mean numbers of zero modes is calculated as
\begin{equation}
	\tilde{N}_\beta(n_p)=\left\langle\beta,t=n_pT\left|\tilde{n}_1\right|\beta,t=n_pT\right\rangle,
\end{equation}
for both GSs with odd and even numbers of fermions, where  $\tilde{n}_1=f_1^\dag f_1$ and $\beta=0,1$. $|\beta,t=n_pT\rangle$ is the time-evolving state initiated with $|\beta\rangle$ which can be numerically calculated by the Lanczos exponentiation.
The physical meaning of $\tilde{N}_\beta(n_p)$ is the survival probability of staying in the states with the zero mode being occupied after quenching .
It can be shown from Eq. (3) that at $V=0$, $\tilde{N}_1-\tilde{N}_0$ is reduced to $\left|\left\langle\psi_1(0)|\psi_1(t)\right\rangle\right|^2$,\cite{Suppl,rajak_survival_2014,sacramento_fate_2014} where $\psi_1(0)$ is the wave function of the edge mode in the Majorana representation and $\psi_1(t)$ is
the time evolution of $\psi_1$ at time $t$ after quenching.
A nonzero value of  $\tilde{N}_1-\tilde{N}_0$ can be viewed as a sign of the topological superconducting phase,
in which there exists survival probability of Majorana edge state. $\tilde{N}_1-\tilde{N}_0 \to 0$ indicates a topologically non-trivial to trivial phase transition, after which the edge states disappear.

\begin{figure}
	\centering
	\includegraphics[width=0.95\linewidth]{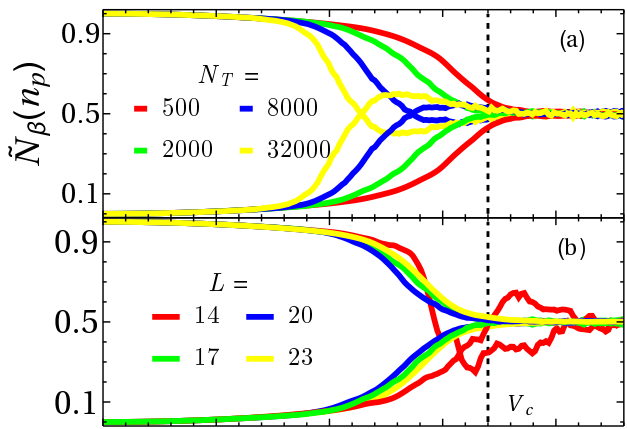}
	\caption{Mean number of zero mode as a function of $V(n_p)$ after slowly quenching from $V=0$.  (a) ``GS'' with different choice of $N_T$ and $L=17$; (b) ``GS'' with different choice of $L$ and $N_T=2000$. $V_M=3.0$ and other parameters are the same as in Fig.\ref{fig:spectrum}}
	\label{fig:quench-dynamics}
\end{figure}

$\tilde{N}_\beta$ as a function of $V(n_p)$ is shown in Fig. \ref{fig:quench-dynamics}. In the initial state, $\tilde{N}_1=1$ and $\tilde{N}_0=0$. After quenching, $\tilde{N}_1$ ($\tilde{N}_0$) continuously decreases (increases) to $\tilde{N}_1(\tilde{N}_0)=0.5$. Two  distinct behaviors in $\tilde{N}_\beta$ can be seen in Fig. \ref{fig:quench-dynamics}.(a). For $N_T =32000$,
$\tilde{N}_0$ first increases to a $N_T$-dependent maximum above 0.5 and then gradually returns to 0.5. The former is due to the tiny avoid crossing between the states with the zero quasi-particle state being occupied and empty for a finite system. Such a non-monotonous change will be suppressed by the Landau-Zener transition for a much faster quenching, e.g., $N_T=500$ and $N_T=2000$ in Fig.~\ref{fig:quench-dynamics}(a), or for a much longer system in Fig.~\ref{fig:quench-dynamics}(b). In these cases, $\tilde{N}_0$ monotonously increases to $\sim 0.5$ around $V_c(n_p) = 1.6\sim 1.8$, which is almost independent either on the chain length, as shown in Fig.~\ref{fig:quench-dynamics}(b) except very short chain, e.g., $L=14$, or on the quenching speed, as shown in Fig.~\ref{fig:quench-dynamics}(a), except that the speed is much faster than the weak adiabatic limit. It then follows that for $\mu_0 =0$
a topological phase transition occurs at $V_c\approx 1.6\sim 1.8$, after which the zero mode fades away.

\emph{Level statistics.-}For $V\neq 0$, the non-integrability of the system can be investigated through level statistics.
The quasi-energy spectrum $\epsilon_\alpha$s is obtained by solving the eigen-value problem, $\hat{U}(T,0)\left|\phi_\alpha\right\rangle=e^{-i\epsilon_\alpha T}\left|\phi_\alpha\right\rangle$, for finite systems with OBCs or PBCs.
We investigate the probability distribution $P(r)$ with $r$ defined as~\cite{dalessio_long-time_2014}
\begin{equation}
	r=\frac{\mathrm{min}(\delta_n,\delta_n+1)}{\mathrm{max}(\delta_n,\delta_n+1)}\in[0,1], \delta_n=\epsilon_{n+1}-\epsilon_{n}.
	\label{eq:5}
\end{equation}
Here, the quasi-energy gaps $\delta_n$ are obtained by first ordering the quasi-energy $\epsilon$ in the interval of $[-\pi,\pi]/T$ and then computing the difference between consecutive values.
$P(r)$ is a frequently used indicator of the integrable-chaos transition and is closely related to the level repulsion~\cite{dalessio_long-time_2014}.
Notice that the analysis of level statistics in the current system is meaningful only in a particular symmetry sector: when different subspaces are mixed, the level repulsion can be missed even if the system is chaotic~\cite{kudo_level_2005,santos_onset_2010,santos_transport_2009}.

\begin{figure}
	\centering
	\includegraphics[width=0.95\linewidth]{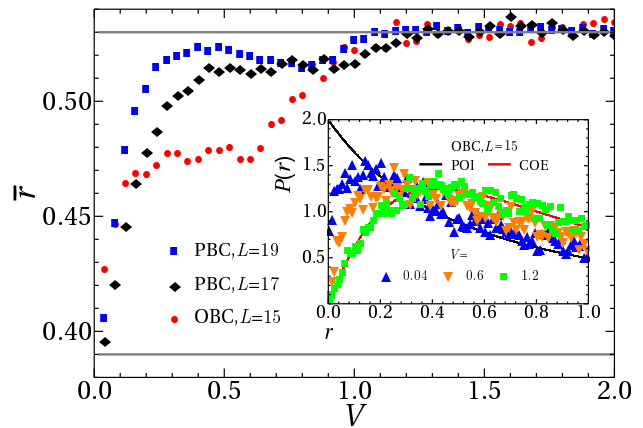}
	\caption{$\bar{r}$ as a function of $V$ for three configurations. The inset shows $P(r)$ for three values of $V$ for a chain of $L=15$ with OBC.
Here $\mu_0 = 0.2$ and the other parameters are the same as in Fig.\ref{fig:spectrum}.}
	\label{fig:levelstatics}
\end{figure}

As shown in the inset of Fig.~\ref{fig:levelstatics}, the system is in a completely chaotic phase for a large interaction, 
e.g., the green squares for $V=1.2$ in the inset of Fig.~\ref{fig:levelstatics},
where a multitude of avoided crossings take place. In this case $P(r)$ (green squares) is close to the distribution of the circular orthogonal ensemble (COE) (red line), for which the vanishing value of $P(r\to 0)$ reflects those level crossings. For smaller interactions, e.g., $V=0.6$ and $V=0.04$, the system is neither regular nor completely chaotic, and $P(r)$ (blue and orange triangles) is distributed between the POI (black line) and COE. There is an increase of the value of $P(r\to 0)$ with $V$ decreased from $V=0.6$ to $V=0.04$, since the avoided crossing gap becomes small. Only in the limit of $V\to 0$, $P(r)$ is in good agreement with the POI.

The mean value of $r$ (denotes by $\bar{r}$) can be used as a signature of the phase transition from regular to chaotic. As shown in Fig.~\ref{fig:levelstatics}, the system is regular  at $V=0$ and $\bar{r}\approx \bar{r}_\textrm{POI}\approx 0.39$. The interaction makes the quantum chaos enhanced, and $\bar{r}$ increases to $\bar{r}_\textrm{COE}\approx 0.53$ for $V_c^\prime \approx 1.2$. Thus, $V_c^\prime \approx 1.2$ is the boundary of semi- to full-chaotic phase transition at $\mu_0 =0.2$.


\begin{figure}
\centering
\includegraphics[width=0.95\linewidth]{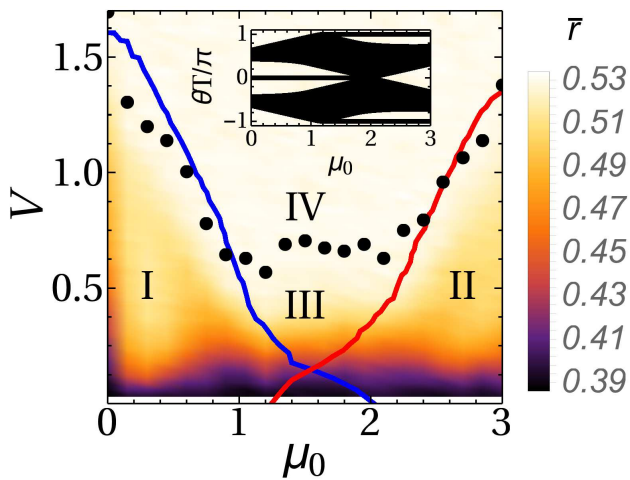}
\caption{Phase diagram in $\mu_0$-$V$ plane. The solid blue (red) line is the boundary below which the Floquet Majorana
zero ($\pi$) mode is preserved. The Black dots constitute the  boundary of transition from semi- to full-chaotic state.
Regions I and II are topologically non-trivial phases, while regions III and IV are trivial. Region IV is completely chaotic
while regions I, II, III are semi-chaotic. The inset shows the one-particle spectrum for $V=0$. Other parameters are the same as in fig. \ref{fig:spectrum}.}
\label{fig:phase}
\end{figure}

\emph{Phase Diagram.-}Using the numerical methods described above, we get the phase diagram in $\mu_0$-$V$ plane, as shown in Fig.~\ref{fig:phase}.
The topological phase boundaries (blue and red lines) are obtained through quench dynamics of ``GS''.
The system is initiated at $\mu_0=V=0$ ($\mu_0=3$ and $ V=0$) such that it is in the topological phase possessing a zero ($\pi$) FMF at one edge.
With slowly varying $V$ and $\mu_0$, $\tilde{N}_1-\tilde{N}_0$ decrease to zero at the quantum critical point of topological phase transition.
The topologically non-trivial/trivial phase boundary in Fig.~\ref{fig:phase} is the contour of $\tilde{N}_1-\tilde{N}_0=0.06$, which is chosen to be very small
but nonzero so that the QCP for the interacting-free case is fixed to $\mu^c_0=2$ ($\mu^c_0=1.2$), as shown in the inset of Fig.~\ref{fig:phase}.
The color in Fig.~\ref{fig:phase} shows $\bar{r}$ for $L=17$ with PBC. Here, $\bar{r}$ has been averaged over $k=0$ with $P=\pm 1$,  $k=1,2,\cdots, L/2-1$, and the subspace with odd/even number of fermions. The dots shows the estimated position at which $\bar{r}$ across $\bar{r}_\textrm{COE}\approx 0.53$,  corresponding to a semi- to full- chaotic phase transition.

It is found that the topologically non-trivial phase is preserved only within a window of $V$
and destroyed as $V$ goes beyond its critical value $V_c(\mu_0)$.
No topologically non-trivial phase exists in the regime of strong interaction,
unlike in the stationary case~\cite{sela_majorana_2011,Hassler2012}. Fig. \ref{fig:phase} shows
three distinct topologically non-trivial phases. At the edge of the chain, they have
only one zero mode, only one $\pi$ mode, and both one zero mode and one $\pi$ mode, respectively,
in region I bounded by blue solid line, region II bounded by red solid line,
and the intersection between them.
The largest $V_c\approx 1.6$ appears at $\mu_0=0$, where the bulk gap with zero mode
in the interacting-free one-particle spectrum is also the largest (see the inset of
Fig.~\ref{fig:phase}), and thus the FMF is most stable.
$V_c$ decreases with decreasing the gap in such an one-particle spectrum.

In the absence of interaction, the excited states are highly degenerate.
Arbitrarily weak interactions will split those degeneracies~\cite{Vasseur2015}.
In the presence of interaction, there exist several regions in the $V$-$\mu_0$ phase diagram,
within which the non-integrable system is topologically non-trivial
and semi-chaotic.
Two regions with different kinds of  competitions between topology and chaos can be seen in Fig.~\ref{fig:phase}.
For $1\lesssim\mu_0\lesssim2.5$, the critical interacting strength, $V_c^\prime$ (black dots),
of the transition from semi-chaotic to full-chaotic states  is apparently larger than $V_c$ and remains
almost constant, $V_c^\prime\approx 0.6$. As a result,
$V_c^\prime \approx 0.6$ is a threshold above which the system is completely chaotic.
For $\mu_0<1$ or $\mu_0>2.5$, however, $V_c^\prime$ is slightly smaller than and almost coincided with $V_c$.
In these regions, $V_c$ or $V_c^\prime$ is larger than $0.6$ and the semi-chaotic region
is enlarged correspondingly. It then follows that topology and chaos can coexist, forming a
new topological state with semi-chaos; and at the same time, they compete with each other,
a full chaotic state requiring a sufficiently strong interaction.
\begin{figure}
	\centering
	\includegraphics[width=0.95\linewidth]{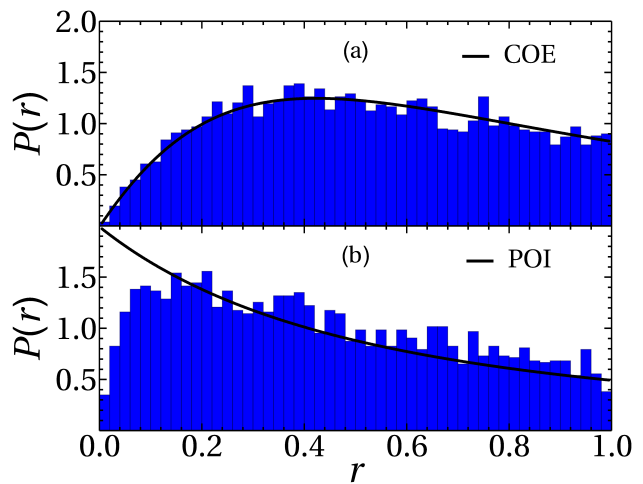}
	\caption{$P(r)$ for the avoided crossing gaps between the states with
		$(\langle \tilde{n}_1 \rangle_i-0.5)\times (\langle \tilde{n}_1 \rangle_{i+1}-0.5)>0$ (a) and $<0$  (b).
		Here $V=0.6$, $\mu_0=0.2$, $L=15$ with OBCs, and other parameters are the same as in Fig.\ref{fig:spectrum} }
	\label{fig:level-same-diff}
\end{figure}

The competition between topology and quantum chaos can be understood
by the following argument. For a system subjected to OBCs,  the topologically non-trivial phase
manifest itself through the existence of localized states at the edges. In the presence
of interaction, even though quasi-particles form many-body states (MBSs) rather than single-particle states,
those MBSs can be roughly classified into two categories: MBSs with the edge mode being
occupied (denoted by OCC) and being empty (EMP). It is expected that the avoided crossing gap between
an OCC and an EMP is generally smaller than that between two OCCs or two EMPs, as can be seen in Fig. 5 below.
 This mechanism is somewhat
similar to the many-body localization, but has a much weaker effect since the bulk states are extended.
In the topologically non-trivial phase in the presence of small $V$,
the OCCs and EMPs can be roughly defined by $\langle \bar{n}_1\rangle_i >0.5$ and $<0.5$, where
$\langle \tilde{n}_1\rangle_i$ is the expected value of $\tilde{n}_1$ for the state with
quasi-energy $\epsilon_i$. At $V=0.6$ and $\mu_0=0.2$, $P(r)$ for the avoided crossing gaps between the states with
$(\langle \tilde{n}_1 \rangle_i-0.5)\times (\langle \tilde{n}_1 \rangle_{i+1}-0.5)>0 $ and $<0$ are shown
in Fig.~\ref{fig:level-same-diff}(a) and (b), respectively. Fig.~\ref{fig:level-same-diff}(a) shows that
if we only consider the gaps between states with the same edge mode occupation,
$P(r)$ will be close to the COE prediction. However, $P(r)$ is still close to the prediction of POI
if we only consider gaps between states with different edge mode occupation (see Fig.~\ref{fig:level-same-diff}(b)).
Thus the system still stays in the semi-chaotic phase if both type of gaps are taken into consideration.

\emph{Conclusion.-}In this work, we have numerically investigated the effect of electron-electron interaction on the topological property of
a 1D periodic driven superconducting chain. We focus on the intermediate-frequency region where neither Magnus-like expansions nor adiabatic
approximations is an appropriate approach. There are two main effects of the interaction: the suppression of the topological superconducting
state and the induction of quantum chaos. A phase diagram has been obtained in the $V$-$\mu_0$ plane. It is found that the topology and quantum
chaos can coexist provided that $V$ is not large enough. They compete with each other, and a sufficiently strong interaction will
lead to a complete chaotic state without any topology.

\end{document}